\documentclass[twocolumn,prb,fleqn]{revtex4}
\usepackage{amsmath}
\usepackage{amsfonts}
\usepackage{amssymb}
\usepackage{graphics}
\usepackage{float}
\usepackage{graphicx}
\usepackage{epstopdf}
\usepackage[compact]{titlesec}
\usepackage{natbib}
\usepackage{threeparttable}

\begin{document}
\title{The evaluation of non-topological components in Berry phase and momentum relaxation time in a gapped 3D topological insulator}
\author{Parijat Sengupta$^{1}$}
\email{parijats@bu.edu}
\author{Gerhard Klimeck$^{2}$}
\author{Enrico Bellotti$^{1}$}
\affiliation{$^{1}$Department of Electrical and Computer Engineering, Boston University, Boston, MA 02215 \\
$^{2}$Dept of Electrical and Computer Engineering, Purdue University \\ 
$^{2}$Network for Computational Nanotechnology, Purdue University, 
West Lafayette, IN, 47907
}

\begin{abstract}
The zero gap surface states of a 3D-topological insulator host Dirac fermions with spin locked to the momentum. The gap-less Dirac fermions exhibit electronic behaviour different from those predicted in conventional materials. While calculations based on a simple linear dispersion can account for observed experimental patterns, a more accurate match is obtained by including higher order $ \overrightarrow{k}$ terms in the Hamiltonian. In this work, in presence of a time reversal symmetry breaking external magnetic field and higher order warping term, alteration to the topologically ordained Berry phase of $ (2n + 1)\pi $, momentum relaxation time, and the magneto-conductivity tensors is established. 
\end{abstract}
\maketitle

\section{Introduction}
Topological insulators (TI) such as Bi$_{2}$Se$_{3}$ and Bi$_{2}$Te$_{3}$ are characterized by surface states which possess distinct characteristics, for instance, a non-trivial Berry phase of $ (2n + 1)\pi $, mass-less Dirac fermions, and a spin helical structure under ideal conditions.~\cite{hasan2010colloquium,qi2011topological} Time reversal symmetry(TRS) breaking perturbations destroy the topological nature of the surface states and introduce additional contributions that have been experimentally established primarily through ARPES data.~\cite{qi2010quantum} Further, the linear dispersion, at points away from the Dirac cone is altered by higher order \textit{k}-terms~\cite{fu2009hexagonal} that in turn modify the topologically ordained behaviour.~\cite{li2015notes,moore2007topological,kane2005z,roy2009topological} An external magnetic field which breaks TRS and higher order \textit{k}-terms that warp the band structure is considered in this work to 1) compute deviations to the Berry phase from the well-known $ (2n + 1)\pi $ and 2) evaluate the momentum relaxation time for electrons scattered on the surface of a 3D-topological insulator.

The paper is outlined as follows: In Section II, an expression for the Berry phase using dispersion and wave functions obtained from the warped and a gapped TI Hamiltonian is calculated which explicitly shows a non-topological component followed by a determination of the momentum relaxation time within a Boltzmann equation framework. The expressions derived in this section are numerically evaluated in Section III. A brief summary of results and their potential implications concludes the paper.

\section{Theory and Model}
The two-dimensional Dirac Hamiltonian is~\cite{zhang2009topological,liu2010model}
\begin{equation}
H_{surf.states} = \hbar v_{f}(\sigma_{x}k_{y} - \sigma_{y}k_{x})
\label{dss}
\end{equation}
where $ v_{f}$ denotes \textit{Fermi}-velocity and $\sigma_{i};{i = x,y} $ are the Pauli matrices. The two-dimensional Dirac Hamiltonian is in principle sufficient to probe the surface states; however, ARPES studies of the \textit{Fermi}-surface at energies significantly far away from the Dirac point reveals a snow-flake like hexagram structure~\cite{alpichshev2010stm,souma2011direct,wang2011observation} that is markedly different from a simple circular Fermi surface observed by application of Eq.~\ref{dss}. This departure from experiment is reconciled by noting that the simple two-dimensional Dirac Hamiltonian fails to account for the underlying crystal symmetries. The deformation of the Fermi surface can be theoretically reproduced if higher order $ k $ terms are incorporated in the Hamiltonian. Since the two-dimensional Dirac Hamiltonian must comply with the C$_{3v}$ point-group and time reversal symmetry, the next higher order terms that must be added are cubic in $ k $. The modified Hamiltonian~\cite{basak2011spin} therefore must look like
\begin{equation}
H(k) = \epsilon_{0}(k) + \hbar v_{f}(\sigma_{x}k_{y} - \sigma_{y}k_{x}) + \dfrac{\lambda}{2}\hbar^{3}\left(k_{+}^{3} + k_{-}^{3}\right)\sigma_{z}
\label{warping}
\end{equation} 
$ \epsilon_{0}(k) $ introduces the particle-hole anisotropy and the cubic terms denote warping. Using Eq.~\ref{warping}, and ignoring particle-hole anisotropy, the surface state spectrum is
\begin{equation}
\epsilon_{\pm}(k) = \epsilon_{0}(k) \pm \sqrt{\hbar v_{f}^{2}k^{2} + \lambda^{2}\hbar^{6}k^{6}cos^{2}(3\theta)}
\label{warpsp} 
\end{equation}
where $ \theta = tan^{-1}\left(\dfrac{k_{y}}{k_{x }}\right) $. The spectrum contains the lowest order correction to the perfect helicity of the Dirac cone predicted in Eq.~\ref{dss}. The $ cos^{2}(3\theta) $ term possesses the symmetry of the C$_{3v}$ point group and the Hamiltonian is evidently time reversal symmetric.

\subsection{Berry phase of gapped Dirac fermions}
The Berry phase of the gapped surface spectrum of a 3D topological insulator was derived in Ref.~\onlinecite{0268-1242-30-4-045004}. A finite band gap can be induced either through the proximity effect of a ferromagnet or an \textit{s}-wave superconductor.~\cite{fu2008superconducting,sengupta2015proximity,lababidi2011microscopic} The final expression for Berry phase is given as
\begin{equation}
\gamma_{\eta} = \pi\left(1 \pm  \dfrac{\Delta_{pro}}{\sqrt{\Delta_{pro}^{2}+ \left( \hbar v_{f}k\right)^{2}}} \right)
\label{finalberry}
\end{equation}
where $ \Delta_{pro} $ is the band gap split on account of proximity effects and $ \eta = \pm 1 $ denotes the helicity. The goal of this section is to examine the influence of warping on the final Berry phase. To begin, the wave functions $ \left(\vert \Psi\left(r;R\right)\rangle\right)$ of the warped Hamiltonian inserted in the Schr{\"o}dinger equation $ H(R)\vert \Psi\left(r;R\right)\rangle = E_{n}(R)\vert \Psi\left(r;R\right)\rangle $ can be written as
\begin{subequations}
\begin{equation}
\Psi_{\eta} = \dfrac{1}{\sqrt{2}}\begin{pmatrix}
\lambda_{\eta}(k)exp(-i\theta) \\
\eta \lambda_{-\eta}(k)
\end{pmatrix}
\label{wfun1}
\end{equation}
where 
\begin{equation}
\lambda_{\eta}(k) = \sqrt{1 \pm \dfrac{\Delta_{pro}^{'}}{\sqrt{\Delta_{pro}^{'2}+ \left( \hbar v_{f}k\right)^{2}}}}
\label{wfun2}
\end{equation}
\begin{equation}
\Delta_{pro}^{'} = \Delta_{pro} + \dfrac{\lambda}{2}\hbar^{3}\left(k_{+}^{3} + k_{-}^{3}\right)\sigma_{z}
\label{mod_delta_pro}
\end{equation}
\end{subequations}
The warping term effectively augments the band gap splitting originally introduced in the topological insulator. The complete Berry phase in \textit{k}-space such that $ \textbf{R} = \textbf{k} $ can be written as
\begin{equation}
\gamma_{n}(C) = i\oint_{C}\langle \Psi\left(r;R\right)\vert \nabla_{R}\vert \Psi\left(r;R\right)\rangle\,dR
\label{bphasecyc}
\end{equation}
Expanding the Berry connection $ A _{\eta} = \langle \Psi\left(r;R\right)\vert \nabla_{R}\vert \Psi\left(r;R\right)\rangle\ $ in Eq.~\ref{bphasecyc} using the wave functions given in Eq.~\ref{wfun1} yields
\begin{equation}
A_{\eta} = \dfrac{i}{2}\begin{pmatrix}
\lambda_{\eta}^{*}exp(i\theta) & \eta \lambda_{-\eta}^{*}
\end{pmatrix}
\begin{pmatrix}
\left( \partial_{k}\lambda_{\eta} - i\lambda_{\eta}\partial_{k}\theta\right)exp(-i\theta) \\
\eta \partial_{k}\lambda_{-\eta}
\end{pmatrix}
\label{bcurv}
\end{equation}
Simplifying the above expression,
\begin{eqnarray}
A_{\eta}(k) & = & \dfrac{i}{2}\left(\lambda_{\eta}^{*}\partial_{k}\lambda_{\eta} -i\vert \lambda_{\eta}\vert^{2}\partial_{k}\theta + \lambda_{-\eta}^{*}\partial_{k}\lambda_{-\eta}\right) \notag \\
& = & \dfrac{1}{2}\vert \lambda_{\eta}(k)\vert^{2}\partial_{k}\theta
\label{bcn2}
\end{eqnarray}
The last expression has been condensed by noting that $ \partial_{k}\lambda_{-\eta} $ evaluates exactly as $ \partial_{k}\lambda_{\eta} $ with sign reversed, therefore taken together they are equal to zero. $ \partial_{k}\lambda_{\eta} $ is worked out. 
\begin{eqnarray}
\partial_{k}\lambda_{\eta}(k) & = & \partial_{k}\sqrt{1 \pm \dfrac{\Delta_{pro}}{\sqrt{\Delta_{pro}^{2}+ \left( \hbar v_{f}k\right)^{2}}}}  \notag \\
                       & = & \eta \dfrac{1}{2\lambda_{\eta}(k)}\partial_{k}{\sqrt{\Delta_{pro}^{2}+ \left( \hbar v_{f}k\right)^{2}}}
\end{eqnarray}
The final expression for Berry connection in matrix notation is therefore
\begin{equation}
A_{\eta} = \dfrac{1}{2}\left(1 \pm \dfrac{\Delta_{pro}}{\sqrt{\Delta_{pro}^{'2}+ \left(\hbar v_{f}k\right)^{2}}}\right)\dfrac{1}{k^{2}}\begin{pmatrix}
-k_{y} \\
k_{x}
\end{pmatrix}   
\label{berryconn}
\end{equation}
In deriving Eq.~\ref{berryconn}, the angular derivatives $ \partial_{k_{x}}\theta(k) = -\dfrac{k_{y}}{k^{2}} $ and $ \partial_{k_{y}}\theta(k) = \dfrac{k_{x}}{k^{2}} $ were used.
Inserting Eq.~\ref{berryconn} in Eq.~\ref{bphasecyc} and integrating over a closed path in \textit{k}-space gives
\begin{equation}
\gamma_{\eta} = \oint_{C} dk_{x} \cdot A_{\eta}(k) + \oint_{C} dk_{y} \cdot A_{\eta}(k)
\label{berryphinit}
\end{equation}
Since the energy contour under the influence of warping is no longer a circle but has a dependence on $ \left(k,\theta\right)$, the usual relations $ k_{x} = kcos(\theta) $ and $ k_{y} = ksin(\theta) $ are modified to $ k_{x} = k(\theta)cos(\theta) $ and $ k_{y} = k(\theta)sin(\theta) $. The angular derivatives are therefore   
\begin{subequations}
\begin{equation}
\dfrac{dk_{x}}{d\theta} = \dfrac{dk}{d\theta}cos\theta - ksin\theta  
\end{equation}
and
\begin{equation}
\dfrac{dk_{y}}{d\theta} = \dfrac{dk}{d\theta}sin\theta + kcos\theta 
\end{equation}
\label{compangder}
The Berry phase with warping using Eq.~\ref{berryphinit} and substituting for angular derivatives from Eq.~\ref{compangder} gives
\end{subequations}
\begin{equation}
\begin{split}
\gamma_{\eta} = \dfrac{1}{2}\oint dk_{x}\left(1 \pm \dfrac{\Delta_{pro}}{\sqrt{\Delta_{pro}^{'2}+ \left(\hbar v_{f}k\right)^{2}}}\right)\dfrac{-ksin\theta}{k^{2}} \\
 + \dfrac{1}{2}\oint dk_{y}\left(1 \pm \dfrac{\Delta_{pro}}{\sqrt{\Delta_{pro}^{'2}+ \left(\hbar v_{f}k\right)^{2}}}\right)\dfrac{kcos\theta}{k^{2}} \\
\gamma_{\eta} = \dfrac{1}{2}\int_0^{2\pi}d\theta\left(1 \pm \dfrac{\Delta + \lambda\hbar^{3}k^{3}cos(3\theta)}{\sqrt{\left(\Delta + \lambda\hbar^{3}k^{3}cos(3\theta)\right)^{2} + \left(\hbar v_{f}k\right)^{2}}}\right)
\label{berrylong}
\end{split}
\end{equation}
An analytic evaluation of the integral is not possible and a numerical solution is presented in Section III. The integral, as evident, consists of the topological contribution of $ \pi $ and the non-topological part shown in Eq.~\ref{nontopberry}. The warping term effectively increases the proximity induced band gap and the non-topological component of the Berry phase.
\begin{equation}
\gamma_{non-top}= \dfrac{\Delta + \lambda\hbar^{3}k^{3}cos(3\theta)}{\sqrt{\left(\Delta + \lambda\hbar^{3}k^{3}cos(3\theta)\right)^{2}  + 
\left(\hbar v_{f}k\right)^{2}}} 
\label{nontopberry}
\end{equation}
The Berry phase induced fictitious magnetic field~\cite{grosso2014solid,xiao2010berry} by computing $ \bigtriangledown \times A_{\eta} $ and setting the warping term to zero is
\begin{equation}
\overrightarrow{B}_{fic} = \mp \dfrac{1}{2}\dfrac{\hbar^{2} v_{f}^{2}\Delta_{pro}}{\left(\sqrt{\Delta_{pro}^{2}+ \left( \hbar v_{f}k\right)^{2}}\right)^{3}}
\label{berb}
\end{equation}
The additional fictitious magnetic field generates a velocity given as $ v_{a} = \dfrac{dk}{dt} \times \overrightarrow{B}_{fic} $. This velocity is transverse to the electric field and gives rise to the intrinsic Hall current. Writing $ \dfrac{dk}{dt} = -\dfrac{e}{\hbar}\bf{E} $, the corresponding Hall current is
\begin{subequations}
\begin{eqnarray}
j_{Hall} = -e\sum_{\sigma}\int_{BZ}\dfrac{dk}{4\pi^{2}}f(k)v_{a} \\
 = -e^{2}\sum_{\sigma}\overrightarrow{E} \times \int_{BZ}\dfrac{dk}{4\pi^{2}}f(k)\overrightarrow{B}_{fic} 
\end{eqnarray}
The Hall conductivity~\cite{laughlin1981quantized,streda1982theory} which is $ \sigma_{xy} = \dfrac{\partial}{\partial E_{y}}\left(j_{Hall,x}\right) $ and evaluating the curl such that $ \dfrac{\partial}{\partial E_{y}}\left(\bf{E} \times \bf{B}\right) = B_{fic,z} $ gives
\begin{equation}
\sigma_{xy} = \dfrac{e^{2}}{\hbar}\int_{BZ}\dfrac{dk}{4\pi^{2}}f(k)B_{fic,z}
\label{anmvel}
\end{equation}
where $ f(k) $ is the electron distribution of the given band.
\end{subequations}
The Berry supported anomalous velocity~\cite{nagaosa2010anomalous,laughlin1983anomalous} in Eq.~\ref{anmvel} gives rise to a orbital magnetization($ \overrightarrow{B}_{fic}$) dependent Hall effect which flows transverse to the electric field. The anomalous velocity ceases to exist for a zero-gap system as can be easily derived by setting $ \Delta $ to zero in Eq.~\ref{berryconn} and evaluating the curl of the Berry vector potential. In presence of a proximity induced band gap, further augmented by warping, the anomalous velocity is finite, a consequence of which is the intrinsic anomalous Hall conductivity. This anomalous Hall conductivity is therefore band gap and warping dependent.

\subsection{Momentum relaxation time}
The reciprocal of the relaxation time by solving the Boltzmann equation is~\cite{ashcroft1976solid,han2012a} 
\begin{equation}
\dfrac{1}{\tau} = \beta\int \dfrac{d^{3}k^{'}}{8\pi^{3}}\delta\left(\varepsilon_{k} - \varepsilon_{k}\right)\vert \chi_{kk^{'}}\vert\left(1 - cos\theta\right)
\label{simbz}
\end{equation}
where $ \chi_{kk^{'}} = \vert\langle\Psi^{'}\vert\Psi\rangle\vert^{2}\zeta\left(s,s^{'}\right) $. The additional spin-scattering factor~\cite{ozturk2014influence} $ \zeta\left(s,s^{'}\right) $ takes in to account the helical spin structure of the TI surface states. If the external magnetic field is sufficiently large, such that the spins are aligned parallel to it, $ \zeta\left(s,s^{'}\right)$ can be set to unity.  $\beta $ is a constant determined from nature of the scattering source.
The wave vector scattering matrix can be computed as follows
\begin{subequations}
\begin{equation}
T(k,k^{'}) =  \vert\dfrac{1}{2}\begin{pmatrix}
\lambda_{+}(k)exp(i\theta) & \lambda_{-}(k)
\end{pmatrix} \begin{pmatrix}
\lambda_{+}(k) \\
\lambda_{-}(k)
\end{pmatrix}\vert^{2}
\end{equation}
where $ \lambda_{\pm} $ is given by Eq.~\ref{wfun2}. Expanding the inner product gives
\begin{equation}
T(k,k^{'}) = \dfrac{1}{4}\left[\left(\lambda_{+}^{2}cos\theta + \lambda_{-}^{2}\right)^{2} + \lambda_{+}^{2}sin\theta\right] 
\label{matelmf}  
\end{equation}
Simplifying Eq.~\ref{matelmf}, one obtains
\begin{equation}
T(k,k^{'}) = \dfrac{\left[2\Delta^{2} + \left(\hbar v_{f}k\right)^{2}\left(1 + cos\theta\right)\right]}{2\left(\Delta^{2} + \left(\hbar v_{f}k\right)^{2}\right)}
\label{matelme} 
\end{equation}
\end{subequations}
In deriving Eq.~\ref{matelme}, it is assumed that scattering takes place between two equi-energetic states such that $ \vert k \vert = \vert k^{'} \vert $. Further, the states are assumed to be of positive helicity (conduction band for a topological insulator) and $ exp(i\theta)$ is set to unity for the initial wave function($\theta = 0$). The relaxation time can now be evaluated by multiplying Eq.~\ref{matelme} by the factor~\cite{mahan2000many} $ (1 - cos\theta) $ and integrating over all angles
\begin{eqnarray}
\dfrac{1}{\tau} &=& \beta\int_{0}^{\pi}T(k,k^{'})\left(1-cos\theta\right)d\theta \notag \\
&=& \dfrac{\pi\beta}{4}\dfrac{\left(\hbar v_{f}k\right)^{2}+ 4\Delta^{2}}{\left(\Delta^{2} + \left(\hbar v_{f}k\right)^{2}\right)} 
\label{antau}
\end{eqnarray}
Setting $ \Delta $ to zero for a pristine 3D-topological insulator, the relaxation time reduces to $ \dfrac{1}{\tau} = \dfrac{\pi\beta}{4} $. The ratio of the relaxation time is therefore
\begin{eqnarray}
\dfrac{\tau_{t}}{\tau_{nt}} &=& \dfrac{\left(\hbar v_{f}k\right)^{2}+ 4\Delta^{2}}{\left(\hbar v_{f}k\right)^{2}+ \Delta^{2}} \notag \\
& = & \dfrac{1 + 4\kappa}{1 + \kappa}
\label{ratio1}
\end{eqnarray}
where $ \tau_{t}\left(\tau_{nt}\right) $ denotes the individual relaxation time for a topological(trivial) insulator and $ \kappa = \dfrac{\Delta^{2}}{\left(\hbar v_{f}k\right)^{2}} $. If $ \kappa \gg 1 $ when a large magnetic field is impressed, the ratio is approximately equal to four. The scattering time for a topological insulator with a band gap is therefore reduced by a factor of four over its zero-gap counterpart.

When the warping term is explicitly included, the band gap $ \Delta $ must be modified to $ \Delta_{warp} = \Delta + \hbar^{3}\lambda cos3\theta k^{3} $. Using $ \Delta_{warp} $, the scattering time expression in Eq.~\ref{antau} changes to
\begin{equation}
\dfrac{1}{\tau} = \alpha \int_{0}^{\pi}\dfrac{2\left(\Delta + \lambda\hbar^{3}cos3\theta k^{3}\right)^{2}\left(1-cos\theta\right)+\left(\hbar v_{f}k\right)^{2}sin^{2}\theta}{2\left\lbrace \left(\Delta + \lambda\hbar^{3}cos3\theta k^{3}\right)^{2}+\left(\hbar v_{f}k\right)^{2}\right\rbrace }d\theta
\label{warptau}
\end{equation}
The integral in Eq.~\ref{warptau} is numerically evaluated in Section III.

The scattering time expression derived in Eq.~\ref{warptau} can be inserted in a Boltzmann equation and solved under the relaxation time approximation to determine the response of the two-dimensional electronic system through the magneto-conductivity tensors $ \sigma_{xx} $ and $ \sigma_{xy} $. The magneto-conductivity tensors are evaluated with an external magnetic field directed along the \textit{z}-axis and an electric field that is confined to the \textit{x-y} plane. The electric current in terms of Boltzmann transport equation per spin is therefore
\begin{equation}
J = \dfrac{e}{4\pi^{2}}\int d^{2}k v_{k}\delta f_{k}
\label{bzcur}
\end{equation}
where $ \delta f_{k} $ is the deviation from the Fermi-Dirac distribution, $ v_{k} = v_{f}\left(cos\theta,sin\theta\right)$ and $ \theta = tan^{-1}\left(\dfrac{k_{y}}{k_{x }}\right) $. $ v_{f} $ is the \textit{Fermi}-velocity. The Boltzmann equation assuming no spatial variation is
\begin{subequations}
\begin{equation}
\dfrac{\partial f(r,k,t)}{\partial t} + \dfrac{dk}{dt}\cdot \dfrac{\partial f(r,k,t)}{\partial k} = 0
\label{sbz1}
\end{equation}
where $ f(r,k,t)$ is the distribution function written as a sum of the equilibrium distribution and deviation under an electric and magnetic field $ f(r,k,t) = f_{0}(r,k,t) + \delta f(r,k,t) $. Since $ \dfrac{dk}{dt} = -q\left(\textbf{E} + \textbf{v} \times \textbf{B}\right) $, Eq.~\ref{sbz1} can be written as
\begin{equation}
-\dfrac{\delta f}{\tau} -q\left(\textbf{E}\cdot v_{k}\dfrac{\partial f}{\partial \varepsilon} + \left(\textbf{v} \times \textbf{B}\right)\cdot \dfrac{\partial \left(\delta f\right)}{\partial k}\right)  = 0
\label{sbz2}
\end{equation}
\end{subequations}
In writing Eq.~\ref{sbz2}, $ \dfrac{\partial f(r,k,t)}{\partial t} $ is approximated as $ -\dfrac{\delta f}{\tau} $ using the relaxation time approximation. \textbf{E} is the electric field on surface of the TI and $\tau $ is carrier relaxation time. 
Solving Eq.~\ref{sbz2} and inserting $ \delta f $ in Eq.~\ref{bzcur}, yields the magneto-conductivity tensors
\begin{subequations}
\begin{equation}
\sigma_{xx} = \dfrac{e^{2}\vert \varepsilon_{f}\vert}{\pi \hbar^{2}}\dfrac{\tau}{1+\omega_{c}^{2}\tau^{2}}
\label{sxx}
\end{equation}
\begin{equation}
\sigma_{xy} = \dfrac{e^{2}\vert \varepsilon_{f}\vert}{\pi \hbar^{2}}\dfrac{\omega_{c}\tau^{2}}{1+\omega_{c}^{2}\tau^{2}}
\label{sxy}
\end{equation}
\end{subequations}
The ratio of the longitudinal and transverse conductivity tensors for a pristine and gapped topological insulator is therefore 
\begin{subequations}
\begin{equation}
\dfrac{\sigma^{nt}_{xx}}{\sigma^{t}_{xx}} = \dfrac{\tau_{nt}\left(1+\omega_{c}^{2}\tau^{2}_{t}\right)}{\tau_{t}\left(1+\omega_{c}^{2}\tau^{2}_{nt}\right)}  
\label{sxxr}
\end{equation}
and
\begin{equation}
\dfrac{\sigma^{nt}_{xy}}{\sigma^{t}_{xy}} = \dfrac{\tau_{nt}^{2}\left(1+\omega_{c}^{2}\tau^{2}_{t}\right)}{\tau_{t}^{2}\left(1+\omega_{c}^{2}\tau^{2}_{nt}\right)}  
\label{sxyr}
\end{equation}
where $ \omega_{c} = e v_{f}^{2}B/\varepsilon $ is the cyclotron frequency for Dirac fermions at a certain energy $ \varepsilon $. 
\end{subequations}

\section{Results}
\begin{figure}
\includegraphics[scale= 1]{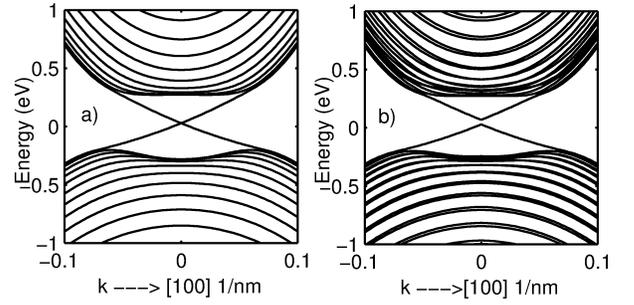}
\caption{Dispersion of a 20.0 $ \mathrm{nm} $ thick Bi$_{2}$Se$_{3}$ topological insulator slab around the Dirac point (Fig.~\ref{fig1}a) at 0.02 $\mathrm{eV}$. The dispersion has a finite band gap (Fig.~\ref{fig1}b) when a ferromagnet is coated on the surfaces. The ferromagnet is assumed to possess an out of plane(\textit{z}-directed) magnetization axis. The magnetization strength is an exchange energy set to 30.0 $\mathrm{meV}$}
\label{fig1}
\end{figure}

The presence of a non-topological component in a gapped 3D-TI (Fig.~\ref{fig1}) is numerically evaluated using expressions derived in Sec II. The surface state of the TI is assumed to possess a finite band gap brought about most commonly through the proximity effect by interfacing with a ferromagnet that has an out-of-plane magnetization component.
\subsection{Numerical evaluation of Berry phase with warping}
\begin{figure}
\includegraphics[scale= 0.9]{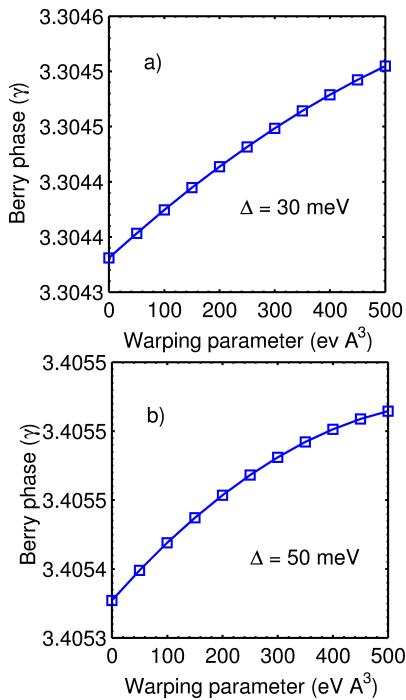}
\caption{The Berry phase ($\gamma$) magnitude plotted as function of the warping strength. $\gamma$ increases with greater warping thus magnifying the non-topological contribution to the standard value of $ \pi $. The overall phase is shown for two values of the band gap indicated on the sub-plots. As warping increases, the band gap contribution diminishes and the curve flattens out.}
\label{fig2}
\end{figure}

The combined influence of warping and a finite band gap on the Berry phase is computed numerically using Eq.~\ref{berrylong}. The warping term $ \hbar^{3}\lambda (\mathrm{meV \AA^{3}}) $ and the band gap $ \Delta $ are varied while $ \vert k \vert $ is held constant at 0.03 $ 1/\AA $. The accumulated Berry phase for a surface electron with positive spin helicity in presence of proximity induced magnetic field and higher order warping terms is shown in Fig.~\ref{fig2}. As the strength of the warping term and band gap increases, the deviation from the topologically determined value of $ \pi $ is more pronounced. The non-topological contribution arising purely on account of warping in absence of a band gap can be computed by setting $ \Delta $ to zero in Eq.~\ref{nontopberry}. Further, if the warping contribution is chosen such that it is equal in magnitude to the surface state energy because of the linear Hamiltonian, the Berry phase expression changes to
\begin{equation}
\gamma_{\eta} = \dfrac{1}{2}\int_0^{2\pi}d\theta\left(1 \pm \dfrac{1}{\sqrt{2}}\right) 
\end{equation}
which integrates to $ 1.292\pi $ for a positive helicity electron. This condition can be fulfilled when $ \hbar v_{f}k = \lambda\hbar^{3}k^{3}cos(3\theta) $. Setting $ cos\left(3\theta\right) $ to 0.5, the warping term~\cite{adroguer2012diffusion} to 200 $ \mathrm{ev{\AA}^{3}}$, and $ v_{f} = 5 \times 10^{5}$ m/s, gives \textit{k} =  0.18 {\AA}. This roughly corresponds to an energy equal to 0.4 $ \mathrm{ev} $ for the surface states. This in a way also determines a cut-off energy beyond which the warped Hamiltonian is the dominant component and outweighs the linear contribution.

A significant manifestation of the increase in Berry phase can be seen as a way to enhance the Berry curvature (Eq.~\ref{berb}) which is the geometric analog of a real magnetic field. Figure~\ref{fig2} shows the dependence of the dimensionless ratio of the Berry curvature to square of the magnetic length $\left(l^{2}_{b} = \dfrac{\hbar}{eB}\right)$ as a function of an external magnetic field. The magnetic length is approximated as 25.0 $ \mathrm{nm}/\sqrt{B}$. The band gap split is calculated using the Zeeman splitting $\left(g\mu_{B}B\right)$, where the \textit{g}-factor is set to 20 for electrons on surface of a TI. The \textit{Fermi}-velocity for surface electrons in Bi$_{2}$Te$_{3}$ is $5 \times 10^{5}$ m/s and the \textit{k}-vector is 0.3 1/{\AA}. The Berry curvature in Eq.~\ref{berb} does not include the warping term, which has been ignored to write a more compact expression.

\begin{figure}
\includegraphics[scale= 0.9]{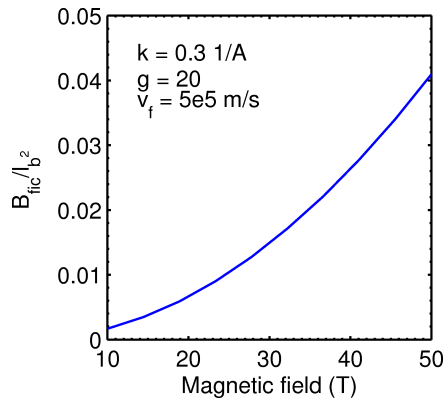}
\caption{The ratio of Berry phase induced orbital magnetic field (Berry curvature) to magnetic length is plotted against an externally applied magnetic field. The Berry curvature is augmented as the magnetic field strength increases. Warping effects are ignored in this calculation. }
\label{fig3}
\end{figure}  
\begin{figure}[h]
\includegraphics[scale= 1]{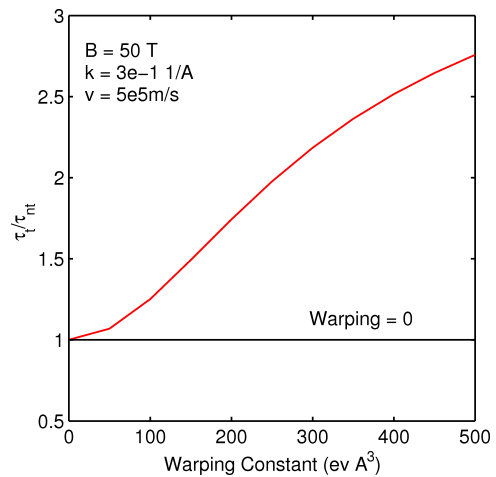}
\caption{The relaxation time is computed for various warping strengths. At zero warping strength, the relaxation times for a finite band gap topological insulator coincide. The relaxation time is significantly altered due to a higher warping of the bands.}
\label{fig4}
\end{figure}
\subsection{Momentum relaxation time}
The ratio of momentum relaxation times for the case of an ungapped topological insulator to a gapped sample is roughly four times if $ \kappa = \dfrac{\Delta^{2}}{\left(\hbar v_{f}k\right)^{2}} $ in Eq.~\ref{ratio1} is significantly larger than unity. The gap $\left(\Delta = g\mu_{B}B \right)$ is computed by assuming an externally impressed magnetic field. Figure~\ref{fig4} shows the relaxation time ratio $ \tau_{t}/\tau_{nt} $ as a function of the warping strength. The corresponding ratio for a magnetic field equal to 50 T without warping is also indicated as a constant. For relatively small values of the magnetic field, the linearly dispersing surface-state energy $ \hbar v_{f}k $ dominates the Zeeman splitting term. For instance, with B = 50 T along the \textit{z}-axis, and a \textit{g}-factor equal to 20, the Zeeman-induced band gap is roughly 0.0579 $\mathrm{eV}$ while the surface energy contribution at \textit{k}-point = 0.3 1/{\AA} is 6.2 $\mathrm{eV}$. The surface energy therefore masks the Zeeman split and brings the ratio of scattering times for a topological and trivial insulator close to unity. For \textit{k}-points sufficiently close to the Dirac point, which is a low energy case, for example, at \textit{k} = 5e-3 1/{\AA}, the ratio changes to 1.71. The overall validity of a surface state energy contribution limited to a linear Hamiltonian holds only when the \textit{k}-point is chosen close to the charge neutral Dirac point at $ \Gamma $. The ratio of scattering times for points in momentum-space farther away from $ \Gamma $ is more correctly represented through a numerical evaluation of Eq.~\ref{warptau}. The warping term usually dominates the band gap split induced by the magnetic field and a marked deviation in the ratio of scattering times from the non-warped case is seen in Fig.~\ref{fig4}. A higher warping strength significantly shifts the ratio curve away from values comparable to unity for cases where the topological insulator is pristine or the higher-order \textit{k} terms have limited contribution.

The ratio of longitudinal and transverse conductivity tensors without warping is evaluated by using Eq.~\ref{sxxr} and Eq.~\ref{sxyr} under a sufficiently strong magnetic field of 10 T at 0.2 $\mathrm{eV} $. The cyclotron frequency is 1.25 GHz; the ratio $ \sigma_{xx}^{nt}/\sigma_{xy}^{nt} $ by approximating $ 1+\omega_{c}^{2}\tau^{2}_{nt} $ as $ \omega_{c}^{2}\tau^{2}_{nt} $ is $ \tau_{t}/\tau_{nt} $. The transverse conductivity under these conditions is almost close to unity. As mentioned above, at points in proximity to $ \Gamma $, such as \textit{k} = 5e-3 1/{\AA}, the longitudinal conductivity tensor ratio is 1.71. The longitudinal conductivity tensor therefore exhibits the same behaviour as the scattering times noted above. 

\section{Conclusion}
The influence of time reversal symmetry(TRS) breaking magnetic field that opens a band gap at the surface and higher order terms defined by a warping of the eigen energy spectrum substantially alter the topological Berry phase of $ \left(2n + 1\right)\pi $ and any phenomenon that depends on the relaxation rate of the surface carriers. The Berry phase from the topologically determined value of $ \pi $ gets an additional component that depends on the band gap split(TRS breaking) and the warping strength. For a given band gap, the Berry phase increases with the warping strength. The altered Berry phase for a finite gap topological insulator described by a warped Hamiltonian also gives rise to anomalous Hall velocity through orbital magentization. The relaxation time within the Boltzmann formalism (assuming spins aligned to the external magnetic field) is also warping dependent and at higher values offsets the magnetic field splitting. More realistic cases of time relaxation expression can be derived for charged impurities with screening and for the Kondo-effect~\cite{kouwenhoven2001revival,schrieffer1967kondo} which involves spin-flip scattering by magnetic impurities. The spin scattering factor which has been chosen as unity to simplify calculations must be set as a scattering angle dependent quantity in a future work.

\begin{acknowledgements}
One of us (PS) thanks late Prof. Gabriele.F. Giuliani from the Dept. of Physics at Purdue University for introducing him to Berry phase and Prof. Avik Ghosh, Department of Electrical Engineering at the University of Virginia, for reading the manuscript and suggestions to improve the paper. PS also gratefully acknowledges the help offered by Yaohua Tan at the Department of Electrical Engineering, Purdue University.
\end{acknowledgements} 

\bibliographystyle{apsrev}
\bibliography{References} 

\begin{thebibliography}{31}
\expandafter\ifx\csname natexlab\endcsname\relax\def\natexlab#1{#1}\fi
\expandafter\ifx\csname bibnamefont\endcsname\relax
  \def\bibnamefont#1{#1}\fi
\expandafter\ifx\csname bibfnamefont\endcsname\relax
  \def\bibfnamefont#1{#1}\fi
\expandafter\ifx\csname citenamefont\endcsname\relax
  \def\citenamefont#1{#1}\fi
\expandafter\ifx\csname url\endcsname\relax
  \def\url#1{\texttt{#1}}\fi
\expandafter\ifx\csname urlprefix\endcsname\relax\def\urlprefix{URL }\fi
\providecommand{\bibinfo}[2]{#2}
\providecommand{\eprint}[2][]{\url{#2}}

\bibitem[{\citenamefont{Hasan and Kane}(2010)}]{hasan2010colloquium}
\bibinfo{author}{\bibfnamefont{M.~Z.} \bibnamefont{Hasan}} \bibnamefont{and}
  \bibinfo{author}{\bibfnamefont{C.~L.} \bibnamefont{Kane}},
  \bibinfo{journal}{Reviews of Modern Physics} \textbf{\bibinfo{volume}{82}},
  \bibinfo{pages}{3045} (\bibinfo{year}{2010}).

\bibitem[{\citenamefont{Qi and Zhang}(2011)}]{qi2011topological}
\bibinfo{author}{\bibfnamefont{X.-L.} \bibnamefont{Qi}} \bibnamefont{and}
  \bibinfo{author}{\bibfnamefont{S.-C.} \bibnamefont{Zhang}},
  \bibinfo{journal}{Reviews of Modern Physics} \textbf{\bibinfo{volume}{83}},
  \bibinfo{pages}{1057} (\bibinfo{year}{2011}).

\bibitem[{\citenamefont{Qi and Zhang}(2010)}]{qi2010quantum}
\bibinfo{author}{\bibfnamefont{X.-L.} \bibnamefont{Qi}} \bibnamefont{and}
  \bibinfo{author}{\bibfnamefont{S.-C.} \bibnamefont{Zhang}},
  \bibinfo{journal}{Physics Today} \textbf{\bibinfo{volume}{63}},
  \bibinfo{pages}{33} (\bibinfo{year}{2010}).

\bibitem[{\citenamefont{Fu}(2009)}]{fu2009hexagonal}
\bibinfo{author}{\bibfnamefont{L.}~\bibnamefont{Fu}},
  \bibinfo{journal}{Physical review letters} \textbf{\bibinfo{volume}{103}},
  \bibinfo{pages}{266801} (\bibinfo{year}{2009}).

\bibitem[{\citenamefont{Li et~al.}(2015)\citenamefont{Li, Kaufmann, and
  Wehefritz-Kaufmann}}]{li2015notes}
\bibinfo{author}{\bibfnamefont{D.}~\bibnamefont{Li}},
  \bibinfo{author}{\bibfnamefont{R.~M.} \bibnamefont{Kaufmann}},
  \bibnamefont{and}
  \bibinfo{author}{\bibfnamefont{B.}~\bibnamefont{Wehefritz-Kaufmann}},
  \bibinfo{journal}{arXiv preprint arXiv:1501.02874}  (\bibinfo{year}{2015}).

\bibitem[{\citenamefont{Moore and Balents}(2007)}]{moore2007topological}
\bibinfo{author}{\bibfnamefont{J.~E.} \bibnamefont{Moore}} \bibnamefont{and}
  \bibinfo{author}{\bibfnamefont{L.}~\bibnamefont{Balents}},
  \bibinfo{journal}{Physical Review B} \textbf{\bibinfo{volume}{75}},
  \bibinfo{pages}{121306} (\bibinfo{year}{2007}).

\bibitem[{\citenamefont{Kane and Mele}(2005)}]{kane2005z}
\bibinfo{author}{\bibfnamefont{C.~L.} \bibnamefont{Kane}} \bibnamefont{and}
  \bibinfo{author}{\bibfnamefont{E.~J.} \bibnamefont{Mele}},
  \bibinfo{journal}{Physical review letters} \textbf{\bibinfo{volume}{95}},
  \bibinfo{pages}{146802} (\bibinfo{year}{2005}).

\bibitem[{\citenamefont{Roy}(2009)}]{roy2009topological}
\bibinfo{author}{\bibfnamefont{R.}~\bibnamefont{Roy}},
  \bibinfo{journal}{Physical Review B} \textbf{\bibinfo{volume}{79}},
  \bibinfo{pages}{195322} (\bibinfo{year}{2009}).

\bibitem[{\citenamefont{Zhang et~al.}(2009)\citenamefont{Zhang, Liu, Qi, Dai,
  Fang, and Zhang}}]{zhang2009topological}
\bibinfo{author}{\bibfnamefont{H.}~\bibnamefont{Zhang}},
  \bibinfo{author}{\bibfnamefont{C.-X.} \bibnamefont{Liu}},
  \bibinfo{author}{\bibfnamefont{X.-L.} \bibnamefont{Qi}},
  \bibinfo{author}{\bibfnamefont{X.}~\bibnamefont{Dai}},
  \bibinfo{author}{\bibfnamefont{Z.}~\bibnamefont{Fang}}, \bibnamefont{and}
  \bibinfo{author}{\bibfnamefont{S.-C.} \bibnamefont{Zhang}},
  \bibinfo{journal}{Nature physics} \textbf{\bibinfo{volume}{5}},
  \bibinfo{pages}{438} (\bibinfo{year}{2009}).

\bibitem[{\citenamefont{Liu et~al.}(2010)\citenamefont{Liu, Qi, Zhang, Dai,
  Fang, and Zhang}}]{liu2010model}
\bibinfo{author}{\bibfnamefont{C.-X.} \bibnamefont{Liu}},
  \bibinfo{author}{\bibfnamefont{X.-L.} \bibnamefont{Qi}},
  \bibinfo{author}{\bibfnamefont{H.}~\bibnamefont{Zhang}},
  \bibinfo{author}{\bibfnamefont{X.}~\bibnamefont{Dai}},
  \bibinfo{author}{\bibfnamefont{Z.}~\bibnamefont{Fang}}, \bibnamefont{and}
  \bibinfo{author}{\bibfnamefont{S.-C.} \bibnamefont{Zhang}},
  \bibinfo{journal}{Physical Review B} \textbf{\bibinfo{volume}{82}},
  \bibinfo{pages}{045122} (\bibinfo{year}{2010}).

\bibitem[{\citenamefont{Alpichshev et~al.}(2010)\citenamefont{Alpichshev,
  Analytis, Chu, Fisher, Chen, Shen, Fang, and Kapitulnik}}]{alpichshev2010stm}
\bibinfo{author}{\bibfnamefont{Z.}~\bibnamefont{Alpichshev}},
  \bibinfo{author}{\bibfnamefont{J.}~\bibnamefont{Analytis}},
  \bibinfo{author}{\bibfnamefont{J.-H.} \bibnamefont{Chu}},
  \bibinfo{author}{\bibfnamefont{I.~R.} \bibnamefont{Fisher}},
  \bibinfo{author}{\bibfnamefont{Y.}~\bibnamefont{Chen}},
  \bibinfo{author}{\bibfnamefont{Z.-X.} \bibnamefont{Shen}},
  \bibinfo{author}{\bibfnamefont{A.}~\bibnamefont{Fang}}, \bibnamefont{and}
  \bibinfo{author}{\bibfnamefont{A.}~\bibnamefont{Kapitulnik}},
  \bibinfo{journal}{Physical review letters} \textbf{\bibinfo{volume}{104}},
  \bibinfo{pages}{016401} (\bibinfo{year}{2010}).

\bibitem[{\citenamefont{Souma et~al.}(2011)\citenamefont{Souma, Kosaka, Sato,
  Komatsu, Takayama, Takahashi, Kriener, Segawa, and Ando}}]{souma2011direct}
\bibinfo{author}{\bibfnamefont{S.}~\bibnamefont{Souma}},
  \bibinfo{author}{\bibfnamefont{K.}~\bibnamefont{Kosaka}},
  \bibinfo{author}{\bibfnamefont{T.}~\bibnamefont{Sato}},
  \bibinfo{author}{\bibfnamefont{M.}~\bibnamefont{Komatsu}},
  \bibinfo{author}{\bibfnamefont{A.}~\bibnamefont{Takayama}},
  \bibinfo{author}{\bibfnamefont{T.}~\bibnamefont{Takahashi}},
  \bibinfo{author}{\bibfnamefont{M.}~\bibnamefont{Kriener}},
  \bibinfo{author}{\bibfnamefont{K.}~\bibnamefont{Segawa}}, \bibnamefont{and}
  \bibinfo{author}{\bibfnamefont{Y.}~\bibnamefont{Ando}},
  \bibinfo{journal}{Physical review letters} \textbf{\bibinfo{volume}{106}},
  \bibinfo{pages}{216803} (\bibinfo{year}{2011}).

\bibitem[{\citenamefont{Wang et~al.}(2011)\citenamefont{Wang, Hsieh, Pilon, Fu,
  Gardner, Lee, and Gedik}}]{wang2011observation}
\bibinfo{author}{\bibfnamefont{Y.}~\bibnamefont{Wang}},
  \bibinfo{author}{\bibfnamefont{D.}~\bibnamefont{Hsieh}},
  \bibinfo{author}{\bibfnamefont{D.}~\bibnamefont{Pilon}},
  \bibinfo{author}{\bibfnamefont{L.}~\bibnamefont{Fu}},
  \bibinfo{author}{\bibfnamefont{D.}~\bibnamefont{Gardner}},
  \bibinfo{author}{\bibfnamefont{Y.}~\bibnamefont{Lee}}, \bibnamefont{and}
  \bibinfo{author}{\bibfnamefont{N.}~\bibnamefont{Gedik}},
  \bibinfo{journal}{Physical review letters} \textbf{\bibinfo{volume}{107}},
  \bibinfo{pages}{207602} (\bibinfo{year}{2011}).

\bibitem[{\citenamefont{Basak et~al.}(2011)\citenamefont{Basak, Lin, Wray, Xu,
  Fu, Hasan, and Bansil}}]{basak2011spin}
\bibinfo{author}{\bibfnamefont{S.}~\bibnamefont{Basak}},
  \bibinfo{author}{\bibfnamefont{H.}~\bibnamefont{Lin}},
  \bibinfo{author}{\bibfnamefont{L.}~\bibnamefont{Wray}},
  \bibinfo{author}{\bibfnamefont{S.-Y.} \bibnamefont{Xu}},
  \bibinfo{author}{\bibfnamefont{L.}~\bibnamefont{Fu}},
  \bibinfo{author}{\bibfnamefont{M.}~\bibnamefont{Hasan}}, \bibnamefont{and}
  \bibinfo{author}{\bibfnamefont{A.}~\bibnamefont{Bansil}},
  \bibinfo{journal}{Physical Review B} \textbf{\bibinfo{volume}{84}},
  \bibinfo{pages}{121401} (\bibinfo{year}{2011}).

\bibitem[{\citenamefont{Sengupta and Klimeck}(2015)}]{0268-1242-30-4-045004}
\bibinfo{author}{\bibfnamefont{P.}~\bibnamefont{Sengupta}} \bibnamefont{and}
  \bibinfo{author}{\bibfnamefont{G.}~\bibnamefont{Klimeck}},
  \bibinfo{journal}{Semiconductor Science and Technology}
  \textbf{\bibinfo{volume}{30}}, \bibinfo{pages}{045004}
  (\bibinfo{year}{2015}).

\bibitem[{\citenamefont{Fu and Kane}(2008)}]{fu2008superconducting}
\bibinfo{author}{\bibfnamefont{L.}~\bibnamefont{Fu}} \bibnamefont{and}
  \bibinfo{author}{\bibfnamefont{C.~L.} \bibnamefont{Kane}},
  \bibinfo{journal}{Physical Review Letters} \textbf{\bibinfo{volume}{100}},
  \bibinfo{pages}{096407} (\bibinfo{year}{2008}).

\bibitem[{\citenamefont{Sengupta et~al.}(2015)\citenamefont{Sengupta, Kubis,
  Tan, and Klimeck}}]{sengupta2015proximity}
\bibinfo{author}{\bibfnamefont{P.}~\bibnamefont{Sengupta}},
  \bibinfo{author}{\bibfnamefont{T.}~\bibnamefont{Kubis}},
  \bibinfo{author}{\bibfnamefont{Y.}~\bibnamefont{Tan}}, \bibnamefont{and}
  \bibinfo{author}{\bibfnamefont{G.}~\bibnamefont{Klimeck}},
  \bibinfo{journal}{Journal of Applied Physics} \textbf{\bibinfo{volume}{117}},
  \bibinfo{pages}{044304} (\bibinfo{year}{2015}).

\bibitem[{\citenamefont{Lababidi and Zhao}(2011)}]{lababidi2011microscopic}
\bibinfo{author}{\bibfnamefont{M.}~\bibnamefont{Lababidi}} \bibnamefont{and}
  \bibinfo{author}{\bibfnamefont{E.}~\bibnamefont{Zhao}},
  \bibinfo{journal}{Physical Review B} \textbf{\bibinfo{volume}{83}},
  \bibinfo{pages}{184511} (\bibinfo{year}{2011}).

\bibitem[{\citenamefont{Grosso}(2014)}]{grosso2014solid}
\bibinfo{author}{\bibfnamefont{G.}~\bibnamefont{Grosso}},
  \emph{\bibinfo{title}{Solid state physics}} (\bibinfo{publisher}{Academic
  Press, an imprint of Elsevier}, \bibinfo{address}{Amsterdam},
  \bibinfo{year}{2014}), ISBN \bibinfo{isbn}{0123850304}.

\bibitem[{\citenamefont{Xiao et~al.}(2010)\citenamefont{Xiao, Chang, and
  Niu}}]{xiao2010berry}
\bibinfo{author}{\bibfnamefont{D.}~\bibnamefont{Xiao}},
  \bibinfo{author}{\bibfnamefont{M.-C.} \bibnamefont{Chang}}, \bibnamefont{and}
  \bibinfo{author}{\bibfnamefont{Q.}~\bibnamefont{Niu}},
  \bibinfo{journal}{Reviews of modern physics} \textbf{\bibinfo{volume}{82}},
  \bibinfo{pages}{1959} (\bibinfo{year}{2010}).

\bibitem[{\citenamefont{Laughlin}(1981)}]{laughlin1981quantized}
\bibinfo{author}{\bibfnamefont{R.~B.} \bibnamefont{Laughlin}},
  \bibinfo{journal}{Physical Review B} \textbf{\bibinfo{volume}{23}},
  \bibinfo{pages}{5632} (\bibinfo{year}{1981}).

\bibitem[{\citenamefont{Streda}(1982)}]{streda1982theory}
\bibinfo{author}{\bibfnamefont{P.}~\bibnamefont{Streda}},
  \bibinfo{journal}{Journal of Physics C: Solid State Physics}
  \textbf{\bibinfo{volume}{15}}, \bibinfo{pages}{L717} (\bibinfo{year}{1982}).

\bibitem[{\citenamefont{Nagaosa et~al.}(2010)\citenamefont{Nagaosa, Sinova,
  Onoda, MacDonald, and Ong}}]{nagaosa2010anomalous}
\bibinfo{author}{\bibfnamefont{N.}~\bibnamefont{Nagaosa}},
  \bibinfo{author}{\bibfnamefont{J.}~\bibnamefont{Sinova}},
  \bibinfo{author}{\bibfnamefont{S.}~\bibnamefont{Onoda}},
  \bibinfo{author}{\bibfnamefont{A.}~\bibnamefont{MacDonald}},
  \bibnamefont{and} \bibinfo{author}{\bibfnamefont{N.}~\bibnamefont{Ong}},
  \bibinfo{journal}{Reviews of modern physics} \textbf{\bibinfo{volume}{82}},
  \bibinfo{pages}{1539} (\bibinfo{year}{2010}).

\bibitem[{\citenamefont{Laughlin}(1983)}]{laughlin1983anomalous}
\bibinfo{author}{\bibfnamefont{R.~B.} \bibnamefont{Laughlin}},
  \bibinfo{journal}{Physical Review Letters} \textbf{\bibinfo{volume}{50}},
  \bibinfo{pages}{1395} (\bibinfo{year}{1983}).

\bibitem[{\citenamefont{Ashcroft and Mermin}(1976)}]{ashcroft1976solid}
\bibinfo{author}{\bibfnamefont{N.}~\bibnamefont{Ashcroft}} \bibnamefont{and}
  \bibinfo{author}{\bibfnamefont{D.}~\bibnamefont{Mermin}},
  \emph{\bibinfo{title}{Solid state physics}} (\bibinfo{publisher}{Holt,
  Rinehart and Winston}, \bibinfo{year}{1976}), ISBN
  \bibinfo{isbn}{0030839939}.

\bibitem[{\citenamefont{Han}(2012)}]{han2012a}
\bibinfo{author}{\bibfnamefont{F.}~\bibnamefont{Han}}, \emph{\bibinfo{title}{A
  modern course in the quantum theory of solids}} (\bibinfo{publisher}{World
  Scientific}, \bibinfo{address}{Singapore Hackensack, N.J. London},
  \bibinfo{year}{2012}), ISBN \bibinfo{isbn}{9814417149}.

\bibitem[{\citenamefont{Ozturk et~al.}(2014)\citenamefont{Ozturk, Filed~III,
  Eo, Wolgast, Sun, and Kurdak}}]{ozturk2014influence}
\bibinfo{author}{\bibfnamefont{T.}~\bibnamefont{Ozturk}},
  \bibinfo{author}{\bibfnamefont{R.~L.} \bibnamefont{Filed~III}},
  \bibinfo{author}{\bibfnamefont{Y.~S.} \bibnamefont{Eo}},
  \bibinfo{author}{\bibfnamefont{S.}~\bibnamefont{Wolgast}},
  \bibinfo{author}{\bibfnamefont{K.}~\bibnamefont{Sun}}, \bibnamefont{and}
  \bibinfo{author}{\bibfnamefont{C.}~\bibnamefont{Kurdak}},
  \bibinfo{journal}{arXiv preprint arXiv:1412.1007}  (\bibinfo{year}{2014}).

\bibitem[{\citenamefont{Mahan}(2000)}]{mahan2000many}
\bibinfo{author}{\bibfnamefont{G.~D.} \bibnamefont{Mahan}},
  \emph{\bibinfo{title}{Many-particle physics}} (\bibinfo{publisher}{Springer
  Science \& Business Media}, \bibinfo{year}{2000}).

\bibitem[{\citenamefont{Adroguer et~al.}(2012)\citenamefont{Adroguer,
  Carpentier, Cayssol, and Orignac}}]{adroguer2012diffusion}
\bibinfo{author}{\bibfnamefont{P.}~\bibnamefont{Adroguer}},
  \bibinfo{author}{\bibfnamefont{D.}~\bibnamefont{Carpentier}},
  \bibinfo{author}{\bibfnamefont{J.}~\bibnamefont{Cayssol}}, \bibnamefont{and}
  \bibinfo{author}{\bibfnamefont{E.}~\bibnamefont{Orignac}},
  \bibinfo{journal}{New Journal of Physics} \textbf{\bibinfo{volume}{14}},
  \bibinfo{pages}{103027} (\bibinfo{year}{2012}).

\bibitem[{\citenamefont{Kouwenhoven and
  Glazman}(2001)}]{kouwenhoven2001revival}
\bibinfo{author}{\bibfnamefont{L.}~\bibnamefont{Kouwenhoven}} \bibnamefont{and}
  \bibinfo{author}{\bibfnamefont{L.}~\bibnamefont{Glazman}},
  \bibinfo{journal}{Physics world} \textbf{\bibinfo{volume}{14}},
  \bibinfo{pages}{33} (\bibinfo{year}{2001}).

\bibitem[{\citenamefont{Schrieffer}(1967)}]{schrieffer1967kondo}
\bibinfo{author}{\bibfnamefont{J.}~\bibnamefont{Schrieffer}},
  \bibinfo{journal}{Journal of Applied Physics} \textbf{\bibinfo{volume}{38}},
  \bibinfo{pages}{1143} (\bibinfo{year}{1967}).

\end{thebibliography}
\end{document}